\documentclass[12pt]{article}

\advance\voffset by -2.0cm \advance\hoffset by -1.25cm
\usepackage{amsmath}
\usepackage{amssymb}
\usepackage{amsthm}
\usepackage{amsfonts}

\numberwithin{equation}{section}

\theoremstyle{definition}

\newcommand{\CC}{\mathbb{C}} 
\newcommand{\ZZ}{\mathbb{Z}} 
\newcommand{\NN}{\mathbb{N}} 
\newcommand{\PP}{\mathbb{P}} 

\DeclareMathOperator{\Sym}{Sym} 

\newcommand{\be}{\begin{equation}}
\newcommand{\ee}{\end{equation}}

\newlength{\oldcolsep}

\begin{document}

\title{Singular Spin Structures and Superstrings}
\author{Marco Matone}\date{}

\maketitle

\begin{center} Dipartimento di Fisica e Astronomia ``G. Galilei'' \\
 Istituto Nazionale di Fisica Nucleare \\
Universit\`a di Padova, Via Marzolo, 8-35131 Padova,
Italy\end{center}

\medskip

\bigskip

\begin{abstract}

There are two main problems in finding the higher genus superstring measure. The first one is that
for $g\geq 5$ the super moduli space is not projected. Furthermore,
the supermeasure is regular for $g\leq 11$, a bound related to the source of singularities
due to the divisor in the moduli space of Riemann surfaces with even spin structure
having holomorphic sections, such a divisor is called the $\theta$-null divisor.
A result of this paper is the characterization of such a divisor. This is done by first extending
the Dirac propagator, that is the Szeg\"o kernel, to the case of an arbitrary number of zero modes, that leads
to a modification of the Fay trisecant
identity, where the determinant of the Dirac propagators is replaced by the product of
two determinants of the Dirac zero modes.
By taking suitable limits of points on the Riemann surface, this {\it holomorphic Fay trisecant identity} leads to identities that
include points dependent rank 3 quadrics in $\mathbb{P}^{g-1}$. Furthermore, integrating over the homological cycles gives relations
for the Riemann period matrix which are satisfied in the presence of Dirac zero modes. Such identities
characterize the $\theta$-null divisor. Finally, we provide the geometrical interpretation of the above points dependent
quadrics and shows, via a new $\theta$-identity, its relation with the Andreotti-Mayer quadric.

\end{abstract}

\newpage

\tableofcontents

\section{Introduction}

Despite the progress in formulating superstring theories, unlike the Polyakov string, there are still many unsolved problems.
In the case of the bosonic string, the Mumford
form on the moduli space of Riemann surfaces is known in terms of theta functions and prime forms.
However, its expression in
terms of theta constants is known only up to genus 4 \cite{Belavin:1986cy}-\cite{Matone:2012wy}. In particular, the genus 4 Polyakov volume form,
conjectured by Belavin and Knizhnik and proved in \cite{Matone:2012wy}, shows that the main obstacle for the extension to arbitrary genus
is strictly related to the problem of characterizing the Jacobian locus, that is the Schottky problem \cite{SamReview}-\cite{Matone:2007wcs}.
In this respect, in \cite{Matone:2012wy} it has been shown
that the higher genus Mumford form is expressed in terms of vector-valued modular forms associated to the Schottky locus, which in turn suggests
that the string partition function is a multiresidue on the Siegel upper half-space. Such investigations strongly suggest that, as observed
by Belavin and Knizhink, any conformally invariant string theory can be expressed in terms of algebraic objects on the moduli space of Riemann
surfaces. This is in the spirit of the Serre GAGA principle. In this respect, it has been shown that
one can introduce modular invariant regularization of string determinants \cite{Matone:2012kk}
and formulate finite string theories, even in 4 dimension, in terms of products of Mumford forms of different weight \cite{Matone:2012rw}.

\noindent
Finding the superstring measure for arbitrary genus $g$ is much harder than in the bosonic case \cite{DHoker:2002hof}-\cite{Grushevsky:2008zm}.
In particular, its explicit form is known only in the case of genus 1 and 2 and
for $g\geq 5$ the super moduli space is not projected \cite{Donagi:2013dua}. Furthermore,
it turns out that the supermeasure is regular for $g\leq 11$, with the bound related to
possible singularities in the locus in the moduli space of supercurves having
even $\theta$-characteristics with nontrivial sections \cite{Felder:2019iqj}.
A related result has been first argued by Witten who suggested
possible problems starting at genus 11 \cite{Witten:2013tpa}. 
His argument is based on the
observation that for an even $\theta$-characteristic with 2 zero-modes there are, in 10 dimension, 20 in all, and since
there are $2g-2$ odd moduli it follows that for $g\geq 11$
there are sufficient picture-changing operators to absorb the fermion zero-modes and this may lead to a singularity.\footnote{I thank E. Witten for
useful comments on this point.}
The critical point is just when the dimension of $H^0(\Sigma; Ber(\Sigma))$ is not $g|0$ and this happens
when there are spin bundles $L_\delta$ such that
\be
h^0(\Sigma_{\rm red}, L_\delta)\neq 0 \ ,
\label{thetanulldivisor}\ee
with $\Sigma_{\rm red}$ the reduced surface of the super Riemann surface. In this case
the isomorphism $Ber_1=\det {\cal V}$, breaks down \cite{Witten:2013tpa}.
The divisor in the moduli space of Riemann surfaces with an even spin structure ${\cal M}_{g,{\rm spin}+}$ corresponding
to an even $\theta$-characteristic with $h^0(\Sigma,L_\delta)\geq 2$ is called the $\theta$-null divisor.

\noindent
One of the main results of the present paper is the characterization, given in Eq.\eqref{danumerare11569}, of such a divisor.
We start by investigating the properties of singular spin bundles, that is the ones associated to $\theta$-characteristics, both odd and even, such that
$\theta[\delta](0)$ and its gradient vanish. In other words, we will
consider the spin bundles $L_\delta$ with
\be
n=h^0(L_\delta)>1 \ .
\ee
We start by showing that by the Riemann vanishing theorem for singular spin structures
one gets properties and relations that characterize holomorphic sections of $H^0(L_\delta)$.
Such relations correspond to a subset of a wider class of holomorphic sections, namely
the ones associated to the singular $\theta$-divisor.

\noindent
As a first step we introduce the Dirac propagator, or Szeg\"o kernel, $S_\delta(z-w)$,
in the presence of an arbitrary number of Dirac zero modes $n$, extending the case $n=1$, reported, for example, in \cite{DHoker:1989cxq}.
We then note that removing
the term $z-w$ from the argument of the $\theta$-function in the modified Sz\"ego kernel gives
\be
\frac{\theta[\delta](\sum_1^n(z_i-w_i))}{\prod_{i,j} E(z_i,w_j)} \ ,
\label{adding}\ee
which is a holomorphic section that,
after adding the prime forms to get a single valued expression, reproduces one side of the Fay trisecant identity \cite{jfayy}. The novelty
is that now this expression is a holomorphic one. Furthermore,
it turns out that the determinant of the Szeg\"o kernel, which is the other side of the trisecant identity,
is now replaced by the product of two determinants of a basis of $H^0(L_\delta)$, that is of the Dirac zero
modes. In this respect, we note that whereas the Fay trisecant formula can be obtained starting from
the Cauchy determinantal formula on the complex plane $\CC$ by
replacing the factors $z_i-w_j$, $z_i-z_j$ and $w_i-w_j$ by the corresponding prime forms and then adding a theta-function to absorb the multivaluedness,
such a formula, that we call {\it holomorphic Fay trisecant
identity} has no counterpart on $\CC$. We will see that such an identity also implies
relations between different spin structures, that include some invariants.

\noindent
By taking suitable limits of points on the Riemann surface, the holomorphic Fay trisecant formula leads to identities that
include points dependent quadrics in $\PP^{g-1}$ of rank 3. Furthermore, integration over the homological cycles leads to relations
for the Riemann period matrix $\tau$ which are satisfied in the presence of Dirac zero modes. These relations characterize the
Jacobian locus when $\tau$ admits singular spin bundles with $n>1$, and, in particular, must be satisfied by the elements of
the $\theta$-null divisor. Finally, we provide the geometrical interpretation of the
rank 3 quadric and shows, via a new $\theta$-identity, its relation with the Andreotti-Mayer quadric \cite{AndreottiMayer}.

\section{Dirac propagator for singular spin structures}

Let us start by introducing basic facts on the Riemann $\theta$-functions. Excellent references are
\cite{jfayy}\cite{FayMAM}.

\noindent
Let $C$ be a compact Riemann surface of genus $g$ and $K_C$ its cotangent bundle. Let $\alpha_i$, and $\beta_i$, $i=1,\ldots,g$, be a symplectic basis of $H_1(C,\ZZ)$ and $\omega_i$, $i=1,\ldots,g$, the basis of $H^0(K_C)$ such that $\oint_{\alpha_i}\omega_j=\delta_{ij}$. Denote by
$\tau_{ij}:=\oint_{\beta_i}\omega_j$ the Riemann period matrix.

\noindent
Let $C_n=\Sym^n(C)$, $n\in\NN_+$, be the space of effective divisors of degree $n$ on $C$ and $J_n(C)$
the principal homogeneous space of linear equivalence classes of divisors of degree $n$ on $C$. Denote
by
\be
J(C):={\CC}^g/L_\tau \ , \qquad L_\tau:={\ZZ}^g +\tau {\ZZ}^g \ ,
\ee
the Jacobian of $C$. Choose an arbitrary point $p_0\in C$ and denote by $I:C\to J(C)$,
$I(p):=(I_1(p),\ldots,I_g(p))$, $I_i(p):=\int_{p_0}^p\omega_i$,
$p\in C$, the Abel-Jacobi map, which is an embedding of $C$ into the Jacobian. Note that
$J(C)$ is identified with $J_0(C)$: each point of $J_0(C)$ can be expressed as $D_2-D_1$ with $D_1$ and $D_2$ effective divisors of the same degree,
this corresponds to $I(D_1-D_2)\in J(C)$, where $I(\sum_i n_i p_i):=\sum_in_iI(p_i)$, $p_i\in C$, $n_i\in\ZZ$.
Note that all the maps $C^g\to C_g\to J(C)$ are surjective.

\noindent
Let us introduce the $\theta$-function with characteristic  $\delta\equiv\big(^{\delta'}_{\delta''}\big)$, with $\delta',\delta''\in \{0,1/2\}^{2g}$
\begin{align}
\theta[\delta](x,\tau)&=\sum_{n\in\ZZ^g}e^{\pi i(n+\delta')\tau(n+\delta'')+2\pi i (n+\delta')(x+\delta'')}\\
&= e^{\pi i \delta' \tau \delta'+2\pi i \delta'(x+\delta'')}\theta(x+\tau \delta' + \delta'',\tau) \ ,
\label{relationchar}\end{align}
$x\in \CC^g$, where $\theta(x,\tau)\equiv\theta\left[^0_0\right](x,\tau)$. The $\theta$-function
has the quasi-periodicity properties
\be
\theta[\delta](x+n+\tau m,\tau)=
e^{-\pi i {m}\tau m-2\pi i{m}x+2\pi i({\delta'}n-{\delta''}m)}\theta[\delta](x,\tau) \ ,
\ee
$m,n\in\ZZ^g$.
The parity of the $\theta$-function, and of the $\theta$-characteristic $\delta$, is the same of the parity of the integer $4\delta'\cdot\delta''$.
There are $2^{2g}$ different characteristics of definite
parity, $2^{g-1}(2^g+1)$
even and $2^{g-1}(2^g-1)$ odd. By Abel Theorem each one of such characteristics determines
the divisor class of a spin bundle $L_\delta\simeq K^{1/2}_C$,
so that we can call them spin structures.
In the following we will frequently denote the Abel-Jacobi map
of points on $C$ by the points themselves, the meaning being clear from the context.

\noindent
Denote by $\Delta$ the vector of Riemann constants. The results of the present investigation are strictly related to the
Riemann Vanishing Theorem: if $z,p_i$ are arbitrary points of $C$, then
\begin{equation}
\theta(z-\sum_1^g p_i+\Delta,\tau) \ ,
\label{rvth}\end{equation}
either vanishes identically or else it has
$g$ zeros at $z=p_1,\ldots,p_g$. This implies that if
$n=h^0(L_\delta)>0$,
then for arbitrary
points $x_i,y_i\in C$, $i=1,\ldots,n$,
\be
\bigg(\theta[\delta](0),\theta[\delta](x_1-y_1),\ldots,\theta[\delta]\bigg(\sum_1^{n-1}(x_i-y_i\bigg)\bigg)=(0,0,\ldots,0) \ .
\label{nullvector}\ee
Recall that the parity of $n$ is the same of the one of $\delta$.
Denote by $\Theta$ the $\theta$-divisor, that is the set of all $e$ such that $\theta(e)=0$, and by
$\Theta_s$ the singular $\theta$-divisor, that is the sublocus of $\Theta$ whose elements are zeros of $\theta$ of order greater than $1$.
By Riemann's Singularity Theorem it follows that the dimension of $\Theta_s$ for $g\geq 4$ is $g-3$ in
the hyperelliptic case and $g-4$ if $C$ is canonical. The curves admitting singular spin structures form a sublocus
of codimension one in the moduli space of genus $g$ canonical curves.

\noindent
Let $\delta$ be a non-singular odd $\theta$-characteristic, i.e. with $n=1$.
We consider the prime form \footnote{Fay denotes the so defined prime form by $E(x,y)$.
Here we choose the notation $E(y,x)$.} \cite{jfayy}
\be
E(y,x)=\frac{\theta[\delta](y-x)}{h_\delta(x)h_\delta(y)} \ , \qquad \forall x,y\in C \ ,
\label{PrimeForm}\end{equation}
where $h_\delta(x)$ is the square root of the holomorphic 1 differential
\be
h_\delta^2(x)=\sum_{1}^{g}\theta_i[\delta](0)\omega_i(x) \ ,
\label{perE}\ee
and $\theta_i[\delta](0)\equiv \partial_{X_i}\theta[\delta](X)_{|X=0}$, $X\in \CC^g$.
$E(y,x)$ is a holomorphic section of a line bundle
on $C\times C$, with the multi-valuedness properties
\be
E(y+\alpha n+\beta m,x)=e^{-\pi i  m \tau m -2\pi i
 m I(x-y)}E(y,x) \ ,
\ee
$m,n\in\ZZ^g$, and such that $E(y,x)=-E(y,x)$.
In particular, it only vanishes if $y=x$, and if $t$ is a local coordinate at $y\in
C$ such that $h_\delta=dt$, then
\be
E(y,x)=\frac{t(y)-t(x)}{\sqrt{dt(y)}\sqrt{dt(x)}}(1+{\cal
O}((t(y)-t(x))^2)) \ .
\ee
\noindent
For any even $\theta$-characteristic with $n=0$ one defines the Szeg$\rm\ddot o$ kernel, that is the Dirac propagator \cite{jfayy}
\be
S_\delta(z,w)=\frac{\theta[\delta](z-w)}{\theta[\delta](0)E(z,w)} \ .
\label{standardSzego}\ee
The first result concerns the Szeg\"o kernel in the case of arbitrary $n$.

\bigskip

\noindent
{\bf Proposition 1.} Let $p_1,\ldots,p_n, q_1,\ldots,q_n$, $n=h^0(L_\delta)$, be pairwise distinct points of $C$.
The corresponding Szeg$\rm\ddot o$ kernel (Dirac propagator) is
\begin{equation}
S_\delta(z,w)=
\frac{\theta[\delta](z-w+\sum_1^{n}(p_i-q_i))}{E(z,w)\theta[\delta](\sum_1^{n}(p_i-q_i))}\prod_1^n\frac{E(z,p_i)E(w,q_i)}{E(z,q_i)E(w,p_i)} \ , \qquad \forall z,w \in C \ .
\label{numerodadare1}\end{equation}

\bigskip

\noindent
The proof is by inspection. One may check that $S_\delta(z,w)$ is single valued with respect to all points and is a meromorphic function
with respect to the $p_i$'s and $q_i$'s.
Furthermore, besides the pole at $z=w$, it has poles also for $z=p_1,\ldots, p_n$ and for $w=q_1,\ldots,q_n$.

\section{Dirac zero modes and the holomorphic Fay trisecant identity}

\noindent
Note that by \eqref{nullvector}
it follows that the ratio
\be
\frac{\theta[\delta](\sum_1^n(z_i-w_i))}{\prod_{i,j} E(z_i,w_j)} \ ,
\label{questalachiave}\ee
has no poles.
More generally, if $e\in J(C)$ is a zero of $\theta$ of order $n$, that is $h^0(e\otimes\Delta)=n$, then
\be
\frac{\theta(\sum_1^n(z_i-w_i)+e)}{\prod_{i,j} E(z_i,w_j)} \ ,
\label{questalachiaveBBB}\ee
has no poles.
By a suitable insertion of prime forms, \eqref{questalachiave} becomes single-valued, preserving holomorphicity, namely
\begin{equation}
F_\delta(\{z_i\};\{w_i\})=\frac{\theta[\delta](\sum_1^n(z_i-w_i)) \prod_{i<j}E(z_i,z_j)E(w_j,w_i)}{\prod_{i,j} E(z_i,w_j)} \ ,
\label{numerodadare1bisse}\end{equation}
which is a holomorphic 1/2 differential with respect to all points $\{z_i\}$ and $\{w_i\}$ and with spin structure $\delta$.
Note that we used $E(w_j,w_i)$ instead of $E(w_i,w_j)$ to avoid factors such as $(-1)^{n(n-1)/2}$. An expression analogous to \eqref{numerodadare1bisse} follows
by inserting the same prime forms in \eqref{questalachiaveBBB}.

\noindent
The above analysis suggests the following theorem that we will prove by using an adaptation of the proof by Fay of his Proposition 2.16 and Corollary 2.18 in
\cite{jfayy}; see also his Lemma 1.1 in \cite{FayDuke} and Theorem 3.1 in \cite{FayMAM}.

\bigskip

\noindent
{\bf Theorem 2.} Let $z_1,\ldots,z_n,w_1,\ldots,w_n$ be pairwise distinct points of $C$ and $e\in J(C)$ a zero of $\theta$ of order $n$. If $\phi_1^+,\ldots,\phi_n^+$ (resp. $\phi_1^-,
\ldots,\phi_n^-$) is a suitable normalized basis for $H^0(e\otimes\Delta)$ (resp. $H^0((-e)\otimes \Delta)$), then
\be
\frac{\theta(\sum_1^n(z_i-w_i)+e) \prod_{i<j}E(z_i,z_j)E(w_j,w_i)}{\prod_{i,j} E(z_i,w_j)}=\det \phi_i^-(z_j)\det \phi_i^+(w_j) \ ,
\label{numerodadare2FAY}\end{equation}
that, in the case of spin bundles $L_\delta$ with $n=h^0(L_\delta)>0$, reads
\be
\frac{\theta[\delta](\sum_1^n(z_i-w_i)) \prod_{i<j}E(z_i,z_j)E(w_j,w_i)}{\prod_{i,j} E(z_i,w_j)}=\det h_{\delta i}(z_j)\det h_{\delta i}(w_j) \ ,
\label{numerodadare2}\end{equation}
with $h_{\delta 1}(x),\ldots, h_{\delta n}(x)$ a suitably normalized basis of $H^0(L_\delta)$.

\bigskip

\noindent
{\it Proof.}
If $Z=z_1+\ldots+z_n$ is a generic divisor of distinct points with $\det \phi_i^-(z_j)\neq 0$,  then, since
the zero set of $\det \phi_i^-(z_j)\neq 0$ is base independent, it follows that there is no section of $H^0((-e)\otimes\Delta)$
vanishing at $Z$, therefore $h^0(\Delta - e - Z) = 0$. On the other hand, by Riemann-Roch theorem
\be
h^0(\Delta - e - Z)-h^0(\Delta+ e + Z)= {\rm deg}\, (\Delta - e - Z)-g+1 \ ,
\ee
so that $h^0(\Delta+e +Z) = n$. Then $h^0(\Delta +e +Z - W) = 0$ for generic $W=w_1+\ldots + w_n$ and thus  $\theta(Z - W + e)$ is not identically vanishing in
$w_1$, say, for generic $w_2,\ldots,w_g$ and generic $Z$. By Riemann's vanishing
theorem $\theta(Z - W + e)$ has a divisor $R$ (in $w_1$) of degree $g$, the unique $R$
with
\be
R \equiv \sum_j z_j -\sum_{j\neq 1}w_j +e+\Delta \in J_g \ .
\ee
The quotient between the left and the right-hand sides of \eqref{numerodadare2FAY} is thus a singlevalued (not identically $0$) function in $w_1$ for generic
fixed $w_2, \ldots , w_n$ and $z_1,\ldots,z_n$, provided $\det \phi_i^-(z_j)\neq 0$.
Now for generic $w_2,\ldots,w_g$,  the zeroes of $\det\phi_i^+(w_j)$ in $w_1$ are at
\be
\sum_{j\neq 1}w_j  + D_1 \equiv \Delta +e \ ,
\ee
where $D_1$ is a divisor of degree $g-n$.
By Riemann's theorem, $R$ is therefore the divisor $\sum_j z_j + D_1$ and thus the ratio between the left- and the right-hand sides of \eqref{numerodadare2FAY}
has no zeroes or poles in $w_1$ and is a non-zero constant in $w_1$, and likewise constant for all $w_j$'s. Therefore, such a ratio is a non-zero constant
$c_1(z_1,\ldots,z_n)$  for all $w_j$'s and generic $Z$.  Repeating the argument for the divisor $-e$ and  with $Z$, $W$ interchanged, this function of the
$z_j$'s is a constant $c_2(w_1,\ldots,w_n)$ for generic $W$ with $\det\phi_i^+(w_j) \neq 0$. Thus $c_1=c_2=c$ is independent of $Z$ and $W$ but dependent on $e$
and the bases that can be taken to be $1$ by choosing them appropriately. Eq.\eqref{numerodadare2} follows trivially by \eqref{numerodadare2FAY}.

\bigskip

\subsection{Determinants of Dirac zero modes and $\theta$-relations}

We now use the above results to derive relations for the Dirac zero modes. Let us first observe
that in the case $n=1$, which is the non-singular odd spin structure, Eq.\eqref{numerodadare2} reduces to
the definition of prime form. This suggests considering the generalization of \eqref{PrimeForm}
using \eqref{numerodadare2}. To this end we take the
limits $w_i\to z_i$, $i=1,\ldots,n$, of $F_\delta(\{z_i\};\{w_i\})$ to get
\be
{\det}^2 h_{\delta i}(z_j)= F_\delta(\{z_i\};\{z_i\}) \ ,
\ee
that is
\be
{\det}^2 h_{\delta i}(z_j)=\frac{1}{n!}\sum_{i_1,\ldots,i_n}^g\theta_{i_1\ldots i_n}[\delta](0)\omega_{i_1}(z_{1})\cdots\omega_{i_n}(z_{n}) \ ,
\label{limititutti}\ee
showing that the right side has double zeros, so that it has a holomorphic square root
\be
\det h_{\delta i}(z_j)=\bigg(\frac{1}{n!}\sum_{i_1,\ldots , i_n}^g\theta_{i_1\ldots i_n}[\delta](0)\omega_{i_1}(z_{1})\cdots\omega_{i_n}(z_{n})\bigg)^{1/2} \ .
\label{limititutti2}\ee

\bigskip

\noindent
{\bf Corollary 3.} If $z_1,\ldots,z_n,w_1,\ldots,w_n$ are pairwise distinct points of $C$, with $n=h^0(L_\delta)>0$, then
\be
F_\delta(\{z_i\};\{w_i\})=(F_\delta(\{z_i\};\{z_i\})F_\delta(\{w_i\};\{w_i\}))^{1/2} \ ,
\label{micidiale}\ee
and
$$
\frac{\theta[\delta](\sum_1^n(z_i-w_i))  \prod_{i<j}E(z_i,z_j)E(w_j,w_i)}{\prod_{i,j} E(z_i,w_j)}
$$
\be
=\frac{1}{n!}\bigg[\bigg(\sum_{i_1,\ldots, i_n}^g\theta_{i_1\ldots i_n}[\delta](0)\omega_{i_1}(z_{i_1})\cdots\omega_{i_n}(z_{i_n})\bigg)\bigg(\sum_{i_1,\ldots,
i_n}^g\theta_{i_1\ldots i_n}[\delta](0)\omega_{i_1}(w_{i_1})\cdots\omega_{i_n}(w_{i_n})\bigg)\bigg]^{1/2} \ .
\label{numerodadare2BISSE}\ee
In the case $n>1$, this implies
\be
\sum_{i_1,\ldots, i_n}^g\theta_{i_1\ldots i_n}[\delta](0)\omega_{i_1}(z)\omega_{i_2}(z)\omega_{i_3}(z_3)\cdots\omega_{i_n}(z_{n})=0 \ ,
\label{notevolerrima}\ee
that follows by setting $z_2=z_1\equiv z$ in \eqref{limititutti2}.
Furthermore, we have the following factorization property
\be
\frac{\theta(\sum_1^n(z_i-w_i)+e)}{\prod_1^n \theta[\nu_i](z_i-w_i)}=g_-(\{z_j\})g_+(\{w_i\}) \ ,
\label{magicfactorization}\ee
with $g_-$ and $g_+$ some sections on $C$, where for any $e\in J(C)$ zero of $\theta$ of order $n$ and with $\{\nu_1,\ldots,\nu_n\}$ any arbitrary set of odd non-singular $\theta$-characteristics.

\bigskip

\noindent
Eq.\eqref{micidiale} is the trivial identity $(\det h_{\delta i}(z_j)\det h_{\delta i}(w_j))^2= {\det}^2 h_{\delta i}(z_j) {\det}^2 h_{\delta i}(w_j)$
and \eqref{numerodadare2}. Eq.\eqref{numerodadare2BISSE} then follows by \eqref{limititutti2} and \eqref{micidiale}. Eq.\eqref{magicfactorization} follows
by replacing the prime forms in \eqref{numerodadare2FAY} by their expression in \eqref{PrimeForm} and then removing all the terms
which are products of sections of $\{z_i\}$ times sections of $\{w_i\}$.
The factorization property \eqref{magicfactorization} is a phenomenon that does not appear on the Riemann sphere. In the context
of string theory, this appears as a loop correction, reminiscent of the cluster property.

\bigskip

\noindent
The prime form is defined in terms of an arbitrary odd non singular theta characteristic, in this respect the
above investigation shows a natural generalization of the prime form, which has the same structure but now
constructed in terms of $\theta[\delta](\sum_1^n(z_i-w_i))$, $n=h^0(L_\delta)>1$ rather than
$\theta[\nu](z-w)$, $h^0(L_\nu)=1$
\be
E_n(\{z_i\},\{w_i\}):=
\frac{\theta[\delta](\sum_1^n(z_i-w_i))\prod_{i<j}E(z_i,z_j)E(w_j,w_i)}{\det h_{\delta i}(z_j)\det h_{\delta i}(w_j)} \ ,
\label{questalachiavenuova}\ee
that, as expected, by \eqref{numerodadare2} coincides with the product of prime forms
\be
E_n(\{z_i\},\{w_i\})=\prod_{i,j} E(z_i,w_j) \ .
\label{generalprimeform}\ee

\bigskip

\noindent
We have the following relations between sections of different spin structures.

\bigskip

\noindent
{\bf Remark 4.} Let $\delta_1$ and $\delta_2$ be two $\theta$-characteristics such that $h^0(L_{\delta_1})=h^0(L_{\delta_2})=n>0$.
For any set of points $z_1,\ldots,z_n,w_1,\ldots,w_n$ in $C$, we have the invariant ratio
\be
\frac{\theta[\delta_1](\sum_1^n(z_i-w_i))}{\det h_{\delta_1 i}(z_j)\det h_{\delta_1 i}(w_j)}=\frac{\theta[\delta_2](\sum_1^n(z_i-w_i))}{\det h_{\delta_2 i}(z_j)\det h_{\delta_2 i}(w_j)} \ .
\label{invariant}\ee
Furthermore, if $\alpha$ and $\delta$ are $\theta$-characteristics such that $h^0(L_\alpha)=0$ and $h^0(L_\delta)=n>0$, then
\be
 \det h_{\delta i}(z_j)\det h_{\delta i}(w_j) = \frac{\theta[\alpha](0)
 \theta[\delta](\sum_1^n(z_i-w_i))}{\theta[\alpha](\sum_1^n(z_i-w_i))}
\det S_\alpha(z_i-w_j) \ ,
\label{differentST}\ee
or, equivalently,
\be
\det (h_\delta S_\alpha^{-1} h_\delta) = \frac{\theta[\alpha](0)
 \theta[\delta](\sum_1^n(z_i-w_i))}{\theta[\alpha](\sum_1^n(z_i-w_i))} \ .
 \label{compact}\ee

\bigskip

\noindent Eq.\eqref{invariant} is an immediate consequence of \eqref{numerodadare2}, whereas
Eq.\eqref{differentST} follows by \eqref{numerodadare2} and the Fay trisecant identity \cite{jfayy}, stating that
if $e\in \CC^g$, with $\theta(e)\neq 0$, then
\begin{equation}
\frac{\theta(\sum_1^m(z_i-w_i)-e)\theta^{m-1}(e) \prod_{i<j}E(z_i,z_j)E(w_j,w_i)}{\prod_{i,j} E(z_i,w_j)}
=\det \left( \frac{\theta(z_i-w_j-e)}{E(z_i,w_j)}\right) \ ,
\label{FayTrisecantIdentity}\end{equation}
for any set of points, $z_1,\ldots,z_m,w_1,\ldots,w_m$ of $C$.

\section{Quadrics in $\PP^{g-1}$ and characterization of the $\theta$-null divisor}

We now show that Theorem 2 leads to rank 3 quadrics in $\PP^{g-1}$ that depend on points of $C$.
In Sect.\ref{geometrical} we will show its relation with the Andreotti-Mayer quadric \cite{AndreottiMayer}
\be
\sum_{i,j}^g\theta_{ij}(e)\omega_{i}(z)\omega_{j}(z)=0 \ ,
\label{AndreottiMayerQ}\ee
$e\in \Theta_{s}$.

\noindent
We start by restricting to the case of spin bundles with $h^0(L_\delta)=2$ and then will consider the case $h^0(L_\delta)>2$. 
This leads to relations that characterize the $\theta$-null divisor. The results will be generalized
to the case of arbitrary $e\in \Theta_{s}$ in Subsect.\ref{generalization}.

\subsection{Two Dirac zero modes}

\noindent
Let us first show that an immediate consequence of Eq.\eqref{micidiale}  of Corollary 3 leads to a quadric that depends on points of $C$. We then will use
such a derivation to provide an alternative derivation
that involves a divisor analysis and that, in Sect.\ref{geometrical}, will lead to the rulings of the
quadric.

\noindent The derivation of the quadric just follows by noticing that Eq.\eqref{micidiale} implies
\be
F_\delta(z,p;z,p)F_\delta(z,q;z,q)=F_\delta^2(z,p;z,q) \ ,
\label{laquadriche}\ee
which is a quadric involving only three holomorphic 1 differentials with respect to $z$.

\bigskip

\noindent
{\bf Corollary 5.} If $h^0(L_\delta)=2$, then there is the following points dependent rank three quadric in $\PP^{g-1}$
\be
\frac{1}{4}\sum_{i,j}^g\theta_{ij}[\delta](0) \omega_i(z)\omega_j(p)\sum_{k,l}^g \theta_{kl}[\delta](0) \omega_k(z)\omega_l(q)=
 \left(\sum_1^g\frac{\theta_i[\delta](p-q) \omega_i(z)}{E(p,q)}\right)^2\ ,
\label{danumerare11}\end{equation}
with $p\neq q$.

\bigskip

\noindent
Note that in the cases $p=z$ or $q=z$ Eq.\eqref{danumerare11} vanishes so that it reduces to
\be
\sum_{i,j}^g\theta_{ij}[\delta](0)\omega_{i}(z)\omega_{j}(z)=0 \ .
\label{notevolerrimabisse}\ee
This is immediate by Eq.\eqref{laquadriche} recalling that by
\eqref{numerodadare2}
\be
F_\delta(\{z_i\};\{w_i\})=\det h_{\delta i}(z_j)\det h_{\delta i}(w_j) \ .
\ee

\bigskip

\noindent
We now provide an alternative proof of the above theorem which is based on a divisor analysis.
Let us first recall that for any $n=h^0(L_\delta)$ one may arbitrarily choose $n-1$ points for each one of the divisors
corresponding to the holomorphic section of $L_\delta$; once the choice is done the remaining $g-n$ points are then uniquely fixed.
Let us first consider one of the two divisors associated to $h^0(L_\delta)=2$.
By \eqref{relationchar} and the Riemann vanishing theorem it follows that if $h^0(L_\delta)=2$, then
\be
\tau\delta'+\delta''= - p-p_2-\ldots-p_{g-1}+\Delta \ ,
\label{deltachar}\ee
where the first point is denoted $p$ instead of $p_1$ to emphasize that, given an arbitrary
$p\in C$, there are always $g-2$ points $p_2(p),\ldots,p_{g-1}(p)\in C$ such that \eqref{deltachar} is satisfied.
Adding the second divisor, we have
\be
\tau\delta'+\delta''=-p-p_2-\ldots - p_{g-1}+\Delta = -q-q_2-\ldots-q_{g-1}+\Delta \ ,
\label{numerodadare5a}\ee
with $p,q\in C$ arbitrary.
In the case $n=2$ Eq.\eqref{numerodadare1bisse} corresponds to the following holomorphic 1/2 differential
with respect to $z,p,w,q\in C$
\be
F_\delta(z,p;w,q) = \frac{\theta[\delta] (z-w+p-q)E(z,p)E(q,w)}{E(z,w)E(z,q)E(p,w)E(p,q)} \ , \qquad \forall p,q,w,z\in C \ .
\label{numerodadare6a}\ee
Recall that $p_2,\ldots,p_{g-1}$ depend only on $p$, whereas $q_2,\ldots,q_{g-1}$ depend
only on $q$. Since the derivative of the $\theta$-function in \eqref{numerodadare6a} with respect to $w$, evaluated at $w=z$ and divided by
$-E(p,q)$, equals $F_\delta(z,p;z,q)$, by Theorem 2 we have
\be
g_\delta(z) :=\det\begin{pmatrix}
 h_{\delta 1}(z) & h_{\delta 2}(z)\\
h_{\delta 1}(p) & h_{\delta 2}(p)
\end{pmatrix}
\det\begin{pmatrix}
     h_{\delta 1}(z) & h_{\delta 2}(z)\\
h_{\delta 1}(q) & h_{\delta 2}(q)
\end{pmatrix}
=\sum_1^g\frac{\theta_i[\delta](p-q) \omega_i(z)}{E(p,q)} \ .
\label{g(z)}\ee

\bigskip

\noindent
Let us show that
\begin{equation}
{\rm div}\, g_\delta(z)= p+\sum_{i=2}^{g-1}p_i + q+\sum_{i=2}^{g-1} q_i  \ .
\label{Divisorofg(z)}\ee
We first note that \eqref{nullvector} and
\eqref{numerodadare5a} imply the following identity
between divisors with respect to $z$
\be
{\rm div}\,\theta[\delta] (z-w+p-q)={\rm div}\,\theta[0] (z-w-q-p_2-\ldots-p_{g-1}+\Delta)=w+q+\sum_{i=2}^{g-1}p_i \ .
\label{boht}\ee
Therefore, the divisor of $g_\delta(z)$ includes $q+\sum_{i=2}^{g-1}p_i$.
On the other hand, \eqref{g(z)} shows that $g_\delta(z)E(p,q)$ is antisymmetric with respect to $p$ and $q$, so that
even $p$ is a zero of $g_\delta(z)$.  It follows that $g_\delta(z)$ is a holomorphic 1 differential with zeros at
$z=q,p,p_2,\ldots,p_{g-1}$, so that, since
it has fixed $g$ zeros, it follows that this is the unique
holomorphic 1 differential with such zeros, and since there are sections of $H^0(L_\delta)$ with divisors $p,p_2,\ldots,p_{g-1}$
and $q,q_2,\ldots,q_{g-1}$, it follows that $g_\delta(z)$ is their product.

\bigskip

\noindent We now introduce two holomorphic 1 bi-differentials. The first one is $1/2$ of the derivative of
$-\sum_1^g\theta_i[\delta](p-q) \omega_i(z)$ in \eqref{g(z)} with respect to
$q$ evaluated at $q=p$, that coincides with $F_\delta(z,p;z,p)$
\be
f_{\delta 1}(z):= {\det}^2 \begin{pmatrix}
 h_{\delta 1}(z) & h_{\delta 2}(z)\\
h_{\delta 1}(p) & h_{\delta 2}(p)
\end{pmatrix} =\frac{1}{2} \sum_{i,j}^g\theta_{ij}[\delta](0) \omega_i(z)\omega_j(p) \ .
\label{f1}\ee
The other holomorphic 1 bi-differential is $F_\delta(z,q;z,q)$
\be
f_{\delta 2}(z):= {\det}^2 \begin{pmatrix}
 h_{\delta 1}(z) & h_{\delta 2}(z)\\
h_{\delta 1}(q) & h_{\delta 2}(q)
\end{pmatrix} =\frac{1}{2} \sum_{i,j}^g\theta_{ij}[\delta](0) \omega_i(z)\omega_j(q) \ .
\label{f2}\ee
Next, note that since $p,p_2,\ldots, p_{g-1}$ are independent of $q$, it follows that the zeros of
$f_{\delta 1}(z)$ include $p,p_2,\ldots,p_{g-1}$.
On the other hand,
$f_{\delta 1}(z)$ is even under
the exchange of $p$ with $z$, so that $p$ is a double zero of $z$. Therefore $f_{\delta 1}(z)$ is a holomorphic 1 differential in $z$ with fixed
$g$ zeros, so that it must be unique up to a multiplicative constant. This means that it is the square of the 1/2 differential
with zeros at $p,p_2,\ldots,p_{g-1}$.

\noindent
Next, changing the role of $p$ and $q$, one sees that $f_{\delta 2}(z)$
has double zeros in $z= q,q_2\ldots,q_{g-1}$, so that is the square of the other zero mode. Comparing the divisors
we get the rank three quadric
\be
f_{\delta 1}(z)f_{\delta 2}(z)=g_{\delta}^2(z) \ ,
\label{414}\ee
that is Eq.\eqref{danumerare11}.

\subsection{Generalization to arbitrary $n$ and characterization of the $\theta$-null divisor}\label{generalization}

The generalization of the points dependent quadric \eqref{danumerare11} to arbitrary $n$ is again a consequence of Theorem 2.

\bigskip

\noindent
{\bf Corollary 6.}  If $z$, $w_j$, $z_k$, $j,k=2,\ldots,n$, are arbitrary but distinct points of $C$, then, for any $n=h^0(L_\delta)>1$ there is the following rank three quadric in $\PP^{g-1}$
$$
F_\delta(z,z_2,\ldots,z_n;z,z_2,\ldots,z_n)F_\delta(z,w_2,\ldots,w_n;z,w_2,\ldots,w_n)
$$
\be
=F_\delta^2(z,z_2,\ldots,z_n;z,w_2,\ldots,w_n) \ ,
\label{quadricaperognin}\ee
that is
$$
\frac{1}{(n!)^2}\sum_{i_1,\ldots ,i_n}^g\theta_{i_1\ldots i_n}[\delta](0) \omega_{i_1}(z)\omega_{i_2}(z_2)\cdots\omega_{i_n}(z_n)
\sum_{j_1,\ldots ,j_n}^g\theta_{j_1\ldots j_n}[\delta](0) \omega_{j_1}(z)\omega_{j_2}(w_2)\cdots\omega_{j_n}(w_n)
$$
\be
 =\left(\sum_{k=1}^g\frac{\theta_k[\delta](\sum_2^n (z_i-w_i))\omega_k(z)\prod_{2\leq i<j} E(z_i,z_j) E(w_j,w_i)}{{\prod_{i=2,j=2}^n E(z_i,w_j)}}\right)^2\ .
\label{danumerare1156}\end{equation}

\bigskip

\noindent
Furthermore, integrating \eqref{danumerare1156} along the canonical basis of homological cycles of $C$, we get
the following relations involving the Riemann period matrix
$$
\frac{1}{(n!)^2}\sum_{i_1,i_m,\ldots ,i_n}^g\theta_{i_1\ldots i_n}[\delta](0) \omega_{i_1}(z)\tau_{i_mk_m}\cdots\tau_{i_n k_n}
\sum_{j_1,j_p,\ldots ,j_n}^g\theta_{j_1\ldots j_n}[\delta](0) \omega_{j_1}(z)\tau_{j_pl_p}\cdots\tau_{j_n l_n}
$$
\be
=\sum_{j,k}^{g}(\delta|k_m,\ldots,k_n;l_p,\ldots,l_n)_{j,k}
\omega_{j}(z)\omega_k(z) \ ,
\label{danumerare11569}\end{equation}
$n>1$, where we introduced the tensor
\begin{align}
&(\delta|k_m,\ldots,k_n;l_p,\ldots,l_n)_{j,k} \nonumber \\  \nonumber \\
& := \oint_{\gamma_m}\oint_{\gamma_p}
\frac{\theta_j[\delta](\sum_2^n (z_r-w_r))\theta_k[\delta](\sum_2^n (z_r-w_r))\prod_{2\leq r<s} E^2(z_r,z_s) E^2(w_s,w_r)}{{\prod_{r=2,s=2}^n E^2(z_r,w_s)}} \ ,
\end{align}
$$
\gamma_m=\alpha_{i_1}\cup \ldots \cup\alpha_{i_{m-1}}\cup \beta_{k_m}\cup\ldots \cup \beta_{k_n} \ , \qquad \gamma_p=\alpha_{i_1}\cup \ldots \cup\alpha_{i_{p-1}}\cup \beta_{l_p}\cup\ldots \cup \beta_{l_n} \ ,
$$
$m,p=1,\ldots, n$, and $j,k=1,\ldots,g$.

\bigskip

\noindent
We note that the relations \eqref{danumerare11569} characterize the period matrices $\tau$ admitting a singular $\theta$-characteristic, and then
the $\theta$-null divisor.

\noindent
The generalization of the above results to the case of any $e\in \Theta_{s}$ is immediate. In particular, we have

\bigskip

\noindent
{\bf Corollary 7.}
If $e\in J(C)$ is a zero of order $n$ of $\theta$ and
$z_1,\ldots,z_n$, $w_1,\ldots,w_n$, are pairwise distinct points of $C$, then
\be
\det \phi_i^-(z_j)\det\phi_i^+(z_j)= \frac{1}{n!}\sum_{i_1,\ldots,i_n}^g\theta_{i_1\ldots i_n}(e)\omega_{i_1}(z_{1})\cdots\omega_{i_n}(z_{n}) \ ,
\label{limitituttiXXX}\ee
and
$$
\frac{\theta(\sum_1^n(z_i-w_i)+e)  \theta(\sum_1^n(z_i-w_i)- e) \prod_{i<j}E^2(z_i,z_j)E^2(w_j,w_i)}{\prod_{i,j} E^2(z_i,w_j)}
$$
\be
=\frac{(-1)^n}{(n!)^2}\bigg(\sum_{i_1,\ldots, i_n}^g\theta_{i_1\ldots i_n}(e)\omega_{i_1}(z_{i_1})\cdots\omega_{i_n}(z_{i_n})\bigg)\bigg(\sum_{i_1,\ldots,
i_n}^g\theta_{i_1\ldots i_n}(e)\omega_{i_1}(w_{i_1})\cdots\omega_{i_n}(w_{i_n})\bigg) \ ,
\label{numerodadare2BISSE4}\ee
that, setting $z_1=w_1\equiv z$, gives the points dependent quadric
$$
\frac{(-1)^n}{(n!)^2}\sum_{i_1,\ldots ,i_n}^g\theta_{i_1\ldots i_n}(e) \omega_{i_1}(z)\omega_{i_2}(z_2)\cdots\omega_{i_n}(z_n)
\sum_{j_1,\ldots ,j_n}^g\theta_{j_1\ldots j_n}(e) \omega_{j_1}(z)\omega_{j_2}(w_2)\cdots\omega_{j_n}(w_n)
$$
\be
 =\frac{\sum_{j,k}^g\theta_j(\sum_2^n (z_i-w_i)+e)\omega_j(z)\theta_k(\sum_2^n (z_i-w_i)-e)\omega_k(z)\prod_{2\leq i<j} E^2(z_i,z_j) E^2(w_j,w_i)}{{\prod_{i=2,j=2}^n E^2(z_i,w_j)}}\ .
\label{danumerare11567789}\end{equation}

\bigskip

\noindent Eq.\eqref{limitituttiXXX} follows by taking the limits $w_i\to z_i$, $i=1,\ldots,g$, of \eqref{numerodadare2FAY}. Eq.\eqref{numerodadare2BISSE4}
follows by the identity
$$
(\det \phi_i^-(z_j) \det \phi_i^+(w_j))(\det \phi_i^-(w_j)\det \phi_i^+(z_j))
$$
\be
=(\det \phi_i^-(z_j) \det \phi_i^+(z_j))(\det \phi_i^-(w_j)\det \phi_i^+(w_j)) \ ,
\ee
and then expressing its left-hand side by the left-hand side of \eqref{numerodadare2FAY}, noticing that
interchanging $z_i$'s and $w_i$'s gives the factor $(-1)^n$,
whereas its right-hand side is replaced by the right-hand side of
\eqref{limitituttiXXX}.

\noindent
We conclude this section noticing that integrating \eqref{danumerare11567789} along the homological cycles
leads to a straightforward generalization of the relations \eqref{danumerare11569}.

\section{Geometrical interpretation of the points dependent quadric}\label{geometrical}

We now adapt to the case of singular points of order 2 the divisor analysis that led to the proof of \eqref{414}, reported also
in \eqref{danumerare11}. The aim is
to show the relation between \eqref{danumerare11} and the Andreotti-Mayer quadric \eqref{AndreottiMayerQ}.
The generalization of \eqref{danumerare11} to the case $H^0(\pm e\otimes\Delta)$ corresponds to \eqref{numerodadare2BISSE4} with $n=2$.

\noindent
Set $X=(X_1,\ldots,X_g)\in\CC^g$, equivalently $[X]\in \PP^{g-1}$. The canonical curve is then $[\omega(C)]:=[\omega(X)]\in\PP^{g-1}$ for all $X=\omega(z)=(\omega_1(z),\ldots,\omega_g(z))\in\CC^g$, $z\in C$.

\noindent
For a divisor $D=p_1+\ldots+p_{g-1}\in C_{g-1}$ with $h^0(D)=2$, the matrix $\omega_i(p_j)$ has rank $g-2$; set
\be
\Sigma_D={\rm span}\, \{\omega(p_1),\ldots,\omega(p_{g-1})\}\subset \CC^g \ .
\ee
We have ${\rm dim}\, \Sigma_D=g-2$, so that $[\Sigma_D]\leftrightarrow \PP^{g-3}$.

\noindent
If $\theta(z)$ vanishes to second order at $z=e\in \CC^g$, that is
$e$ is a singular point of order two on $\Theta$, then $Q(X):=
\sum_{i,j}\theta_{ij}(e)X_iX_j=0$ is a quadric in $\PP^{g-1}$ containing $[\omega(C)]$, that is Eq.\eqref{AndreottiMayerQ}.
If $e\equiv D-\Delta\in J_0(C)$, then
\be
X\in\Sigma_D \; \Longrightarrow \; Q(X)=0 \ ,
\ee
that is
\be
\sum_{i,j}\theta_{ij}(e)\omega_i(p_\alpha)\omega_j(p_\beta)=0 \ ,
\label{SingularLocus}\ee
for all $\alpha$ and $\beta$ in $[1,g-1]$, which is
Theorem 3.6 of \cite{Matone:2007wcs} in the case $n=2$.

\noindent
$\Sigma_D$ depends on the $\PP^1$-family of $D\in C_{g-1}$, all
giving rise to the same $e\in J_0(C)$. For any $p,q\in C$ set
\be
\begin{cases}
\Delta+e\equiv D_p^+\equiv D_q^+ \ , & D_p^+=p+\xi_p \ ,  \quad D_q^+=q+\xi_q \ , \\
\Delta-e\equiv D_p^-\equiv D_q^- \ , & D_p^-=p+\eta_p \ , \quad D_q^-=q+\eta_q \ ,
\end{cases}
\ee
for divisors $\xi_*$, $\eta_*$ of degree $g-2$ depending on $*$.

\noindent
Setting
\be
K_{p,q}^\pm (X)=\sum_i\theta_i(p-q\pm e) E(p,q)^{-1} X_i \ ,
\ee
\be
H_p(X)=\sum_{i,j}\theta_{ij}(e) X_i\omega_j(p) \ ,
\ee
we have $K_{p,p}^\pm=H_p$, $K_{p,q}^- =K_{q,p}^+$ and
\be
K_{p,q}^+(\omega(z))= \det
\begin{pmatrix}
 \phi_{1}^-(z) & \phi_2^-(z)\\
\phi_1^-(p) & \phi_2^-(p)
\end{pmatrix}
\det
\begin{pmatrix}
 \phi_{1}^+(z) & \phi_2^+(z)\\
\phi_1^+(q) & \phi_2^+(q)
\end{pmatrix} = K_{q,p}^-(\omega(z)) \ .
\ee
\be
H_p(\omega(z))= \det
\begin{pmatrix}
 \phi_{1}^-(z) & \phi_2^-(z)\\
\phi_1^-(p) & \phi_2^-(p)
\end{pmatrix}
\det
\begin{pmatrix}
 \phi_{1}^+(z) & \phi_2^+(z)\\
\phi_1^+(p) & \phi_2^+(p)
\end{pmatrix} \ ,
\ee
These hyperplanes intersect $[\omega(C)]\in \PP^{g-1}$ at
\be
\begin{cases}
{\rm div}\, H_p(\omega(z))=2p + \xi_p+\eta_p \ , & {\rm div}\, H_q(\omega(z))=2q + \xi_q+\eta_q \ , \\
{\rm div}\, K_{p,q}^+(\omega(z))=p + q+\xi_p+\eta_p \ , & {\rm div}\, K_{p,q}^-(\omega(z))=p+q + \xi_p+\eta_q \ ,
\end{cases}
\ee
and if $(Q)={\rm div}\, Q(X)\subset \PP^{g-1}$, the rulings of $Q=0$ by the $\Sigma_* \leftrightarrow \PP^{g-3}$'s are
\be
\begin{cases}
\Sigma_{D_p^+}\cup\Sigma_{D_p^-}\subset (H_p)\cap (Q) \ , &
\Sigma_{D_q^+}\cup\Sigma_{D_q^-}\subset (H_q)\cap (Q) \ , \\
\Sigma_{D_p^-}\cup\Sigma_{D_q^+}\subset (K_{p,q}^+)\cap (Q) \ , &
\Sigma_{D_p^+}\cup\Sigma_{D_q^-}\subset (K_{p,q}^-)\cap (Q) \ .
\end{cases}
\ee
That $Q(X)=0$ is a rank $\leq 4$-quadric can be expressed by the following relation.

\bigskip

\noindent
{\bf Theorem 8.}
\be
H_p(X)H_q(X)-K_{p,q}^+(X)K_{p,q}^-(X)=c_{p,q} Q(X) \ ,
\label{seguedatrisecant}\ee
for all $X\in \CC^g$, and constant
\be
c_{p,q}=\frac{1}{2}\sum_{\alpha,\beta}\theta_{\alpha \beta}(e)\omega_\alpha(p)\omega_\beta(q)\neq 0 \ ,
\ee
for generic $p$ and $q$.

\bigskip

\noindent Eq.\eqref{seguedatrisecant} follows by the
Fay trisecant identity for $n=2$
$$
\theta(x-p-e)\theta(y-q-e)E(x,q)E(p,y)+\theta(x-q-e)\theta(y-p-e)E(x,p)E(y,q)
$$
\be
=\theta(x+y-p-q-e)\theta(e)E(x,y)E(p,q) \ .
\ee
In particular, differentiating this identity with respect to $e_m$ and $e_n$, evaluated at the singular $e$, and then setting $x=p$, $y=q$,
one gets
$$
-[\theta_m(p-q-e)\theta_n(p-q+e)+\theta_n(p-q-e)\theta_m(p-q+e)]E^{-2}(p,q)
$$
\be
=\sum_{\alpha,\beta}(\theta_{mn}(e)
\theta_{\alpha\beta}(e)-\theta_{m\alpha}(e)\theta_{n\beta}(e)-\theta_{n\alpha}(e)\theta_{m\beta}(e))\omega_\alpha(p)\omega_\beta(q) \ .
\ee
Interesting relations, similar
to \eqref{seguedatrisecant}, follow for the tangent cone
\be
\sum_{i_1,\ldots,i_n}\theta_{i_1\ldots i_n}(e)X_{i_1}\cdots X_{i_n} \ , \qquad n\geq 3 \ ,
\ee
for $X=(X_1, \ldots, X_g)\in \CC^g$ at singular $e\in\Theta$.

\bigskip

\noindent
{\bf Acknowledgments.} It is a pleasure to thank
John Fay and  Roberto Volpato for important suggestions and remarks and Samuel Grushevsky
and Edward Witten for key comments.
I also  gratefully acknowledge support from the Simons Center for Geometry and Physics, Stony Brook University at which the final part of this paper was performed.

\end{document}